\RequirePackage{fixltx2e}
\documentclass[prl,twocolumn,amsmath,amssymb,amsfonts,longbibliography,superscriptaddress]{revtex4-1}
\usepackage{tikz}
\usepackage{tikz-cd}
\usepackage{amsmath}
\usepackage{times}
\usepackage[varg]{txfonts}
\usepackage{hyperref}
\usepackage{color}
\usepackage{pdfpages}

\usepackage{mathrsfs}
\usepackage{color}
\usepackage{graphicx}
\usepackage[percent]{overpic}
\usepackage{subcaption}
\usepackage{bm}
\usepackage{fixltx2e}
\usepackage{natbib}

\newcommand{\bra}[1]{\left \langle #1\right|} 
 
\newcommand{\ketp}[1]{|#1\big>} 
 
\newcommand{\ket}[1]{\left| #1  \right \rangle}

\newcommand{\avg}[1]{\langle #1 \rangle}

\newcommand{\h}[1]{{#1}^{\dagger}}

\newcommand{\meV}{{\ \rm meV}}

\newcommand{\K}{{\ \rm K}}
\newcommand{\mK}{{\ \rm mK}}

\newcommand{\tsup}[1]{\textsuperscript{#1}}
\newcommand{\tsub}[1]{\textsubscript{#1}}

\newcommand{\abo}[2]{\text{#1}\tsub{2}\text{#2}\tsub{2}\text{O}\tsub{7}}
\newcommand{\rto}[1]{\abo{#1}{Ti}}
\newcommand{\dbo}[1]{\abo{Dy}{#1}}

\newcommand{\dto}{\abo{Dy}{Ti}}
\newcommand{\dso}{\abo{Dy}{Sn}}
\newcommand{\hso}{\abo{Ho}{Sn}}
\newcommand{\dgo}{\abo{Dy}{Ge}}
\newcommand{\hgo}{\abo{Ho}{Ge}}

\newcommand{\hto}[1]{Ho\tsub{2}Ti\tsub{2}O\tsub{7}}

\newcommand{\rth}[1]{R\tsup{3+}}
\newcommand{\dth}{Dy\tsup{3+}}
\newcommand{\hth}{Ho\tsup{3+}}

\newcommand{\exJ}{J}
\newcommand{\exD}{D}
\newcommand{\exDp}{\mathscr{D}}
\newcommand{\mJ}{{\rm J}}
\newcommand{\mS}{S}
\newcommand{\tamJ}{\rm J}
\newcommand{\termLS}{{}^{2 {\rm S}+1} {\rm L}}
\newcommand{\termLSJ}{\termLS_{\tamJ}}
\newcommand{\muB}{\mu_{\rm B}}
\newcommand{\DeltaOx}{\Delta_{pf}}
\newcommand{\DeltaCF}{\Delta}
\newcommand{\ssection}[1]{\emph{#1:}}
\newcommand{\dgs}{{}^{\rm 6} {\rm H}_{\rm 15/2}}
\newcommand{\eV}{\ {\rm eV}}
\newcommand{\exJtr}{\exJ_{\pm}}

\newcommand{\EE}{effective exchange}
\newcommand{\suppinfo}{Supplemental Material}

\definecolor{acolor1}{RGB}{68,34,221}
\definecolor{acolor2}{RGB}{28,181,0}
\definecolor{acolor3}{RGB}{238,34,34}

\hypersetup{colorlinks=true,linkcolor=acolor1,citecolor=acolor1,urlcolor=acolor1} 

\begin{document}
 
\title{How quantum are classical spin ices?}
\author{Jeffrey G. Rau}
\affiliation{Department of Physics and Astronomy, University of
Waterloo, Ontario, N2L 3G1, Canada} 
\author{Michel J. P. Gingras} 
\affiliation{Department of Physics and Astronomy,
University of Waterloo, Ontario, N2L 3G1, Canada}
\affiliation{Perimeter Institute for Theoretical Physics, Waterloo,
Ontario, N2L 2Y5, Canada} 
\affiliation{Canadian Institute for Advanced
Research, 180 Dundas Street West, Suite 1400, Toronto, ON, M5G 1Z8,
Canada}
\begin{abstract}
The pyrochlore spin ice compounds \dto{} and \hto{} are well described
by classical Ising models down to very low temperatures. Given the
empirical success of this description, the question of the importance
of quantum effects in these materials has been mostly ignored. We show
that the common wisdom that the strictly Ising moments of \dth{} and
\hth{} imply strictly Ising interactions is too na\"ive; a more 
complex argument is needed to explain the close agreement
between theory and experiment.  From a microscopic picture of the
interactions in rare-earth oxides, we show that the high-rank
multipolar interactions needed to induce quantum effects are generated
only very weakly by super-exchange. Using this framework, we formulate
an estimate of the scale of quantum effects in \hto{} and \dto{},
finding it to be well below experimentally relevant temperatures. We
discuss the implications of these results for realizing quantum spin
ice in other materials.
\end{abstract}

\maketitle
\ssection{Introduction} 
Spin ice has proven to be one of the more fruitful marriages of
theoretical and experimental condensed matter physics
\cite{Bramwell_Science,Gingras_Springer,GGG_RMP,Castelnovo_ARCMP,Gingras_RPP}.
This cooperative paramagnetic \cite{Villain} phase is a magnetic
analogue of common water ice \cite{Bramwell_Science,Gingras_Springer},
with proton displacements mapped to magnetic moments pointing in or
out of the corner-shared tetrahedra of the pyrochlore lattice
\cite{Harris_PRL}.  Spin ice displays an exponential number of
low-energy states, and thus an associated extensive residual entropy
\cite{Ramirez_Nature,Cornelius_PRB}.  This manifold is characterized
by the ice rules specifying that on each tetrahedron two spins must
point in, and two must point out.  Generically referred to as a
``Coulomb phase'' \cite{Castelnovo_Nature,Henley_ARCMP}, the spin ice
state harbors a rich phenomenology such as bow-tie shaped
singularities (pinch points) in the magnetic structure factor
\cite{Henley_ARCMP,Isakov_pinch_points} and gapped low-energy
excitations which provide a condensed matter realization of ``magnetic
monopoles'' \cite{Castelnovo_Nature,Castelnovo_ARCMP}.

The two textbook materials that realize this spin ice physics are the
pyrochlore rare-earth titanates \cite{GGG_RMP} \hto{}
\cite{Harris_PRL} and \dto{} \cite{Ramirez_Nature}.  Significant
evidence has accumulated that their magnetic and thermodynamic
properties
\cite{denhertog,Ruff_DTO,Castelnovo_Nature,bramwell2001spin} can be
quantitatively described by a classical model that includes
nearest-neighbor exchange and long-range magnetostatic dipole-dipole
interactions, both of \emph{purely} Ising type. Amendments
incorporating Ising interactions beyond nearest-neighbors have been
proposed to account for various fine details in the thermodynamic and
magnetic behavior of \dto{} \cite{Ruff_DTO,Yavorski_DTO,Tabata}.  The
combination of Ising exchanges and long-range dipolar interactions is
expected to lift the degeneracy of the ice manifold and release the
residual entropy at very low temperatures
\cite{Gingras_CJP,melko2001long,Melko_JPCM,Isakov_rules}.
Experimentally, such a transition has yet to be observed
\cite{fukazawa2002magnetic}, with further significant deviations
\cite{Pomaranski} from predictions raising questions to the
completeness of the classical dipolar Ising spin-ice model
\cite{Yavorski_DTO}.  In particular, one could speculate that the
observed rise of the magnetic specific heat below $\sim 500$ mK
\cite{Pomaranski} is an indication that quantum effects are becoming
important \cite{shannon2013magnetic,mcclarty2014quantum}.

This classical Ising description
\cite{siddharthan1999ising,denhertog,bramwell2001spin} of the
interactions has framed the theoretical and experimental perspectives
\cite{Bramwell_Science,Gingras_Springer,GGG_RMP,Castelnovo_ARCMP,Gingras_RPP}
on these materials dating back to their initial discovery
\cite{Harris_PRL,Ramirez_Nature}.  Beyond its empirical successes,
this mind-set
\cite{siddharthan1999ising,denhertog,Gingras_CJP,melko2001long,bramwell2001spin,Melko_JPCM,Isakov_rules,Ruff_DTO,Yavorski_DTO,Tabata}
is rooted in the \emph{single-ion} magnetic properties of Dy$^{3+}$
\cite{flood1974magnetization,fukazawa2002magnetic,
matsuhira2002low,ke2008magnetothermodynamics} and Ho$^{3+}$
\cite{siddharthan1999ising,rosenkranz2000crystal,petrenko2003magnetization,hallas2012statics},
specifically in the Ising nature of the crystal electric field (CEF)
ground doublet.  A consequence of this strict Ising anisotropy is that the
transverse components of the total angular momentum, $\mJ^{\pm}$,
\emph{vanish} between states of the CEF doublet.  This is sufficient
to explain the Ising form of any bilinear, anisotropic exchange
interactions $\sim \mJ^{\mu}_i K^{\mu\nu}_{ij} \mJ^{\nu}_j$, including
the long-range dipolar interactions \cite{Gingras_Springer}.  The theoretical basis of the
classical Ising description of spin ices then appears to rest on the
implicit assumption that interactions between the rare-earth ions
takes such a bilinear form. This common assumption is in fact
\emph{incorrect}; due the large spin-orbit coupling in rare-earth
ions, strong multipolar interactions between the $\mJ^{\mu}_i$ momenta
are generated when the microscopic super-exchange is downfolded into
the free-ion $\termLSJ$ manifold \cite{santini2009multipolar}.
Indeed, this argument has been invoked \cite{onoda2011quantum} to show
that quadrupolar interactions may not be negligible in Pr-based
pyrochlore oxides with strictly Ising moments, such as \abo{Pr}{Sn}
\cite{zhou2008dynamic} and \abo{Pr}{Zr} \cite{kimura2013quantum}.
Further compounding the problem is the possibility that even moderate
transverse couplings could preempt the ordering expected from the
dipolar interactions. The existence of other spin ice compounds, such
as \hso{} \cite{HSO,matsuhira2002low}, \dso{}
\cite{DSO,matsuhira2002low}, \hgo{} \cite{hallas2012statics,DHGO} and
\dso{} \cite{ke2008magnetothermodynamics,DHGO} makes an explanation
through some accidental fine-tuning unlikely. An unanswered puzzle
thus lies in the identification of \dto{} and \hto{} as spin ices: Why
are these compounds so well described by a classical Ising model down
to low temperatures?  What principles, if any, constrain the scale of
the multipolar interactions and suppress significant quantum
fluctuations?
 
In this letter, we explain why quantum effects in Dy- and Ho-based
spin ice materials are small, providing approximate upper bounds on their
size.  First, we outline possible sources of exchange interactions and
show that only magnetic dipole-dipole (MDD) and super-exchange between
the rare-earth ions are potentially non-negligible at experimentally
relevant temperatures. As the MDD interaction is purely Ising when
projected into the ground CEF doublet, we focus on
super-exchange. Using a microscopic model of the oxygen charge
transfer processes, we find that the resulting multipolar inter-ionic
couplings arising from super-exchange are strongly suppressed beyond rank
seven
\footnote{
While not explicitly stated in the literature, this result is implicit
in many earlier works on super-exchange in rare-earths, see for
example Refs.
[\onlinecite{elliott1968orbital,santini2009multipolar,onoda2011quantum}]. Very
recent work by \citet{iwahara2015exchange} has also explicitly pointed
this out.
}.
Because of the composition of the ground doublet, operators of
rank seven or less can only connect its sub-leading spectral
components and are thus highly suppressed. From
experimental constraints on the size of these sub-leading components,
we formulate estimates of the magnitude of the transverse (non-Ising)
couplings in the low-energy theory, estimating them to be
roughly two orders of magnitude smaller than the nearest-neighbor
Ising coupling.  To make this more concrete, we carry out a
model calculation of the super-exchange interaction in \dto{} and
\hto{}. This calculation corroborates our argument and
provides explicit estimates for the size of quantum effects in the
\hto{} and \dto{} spin ices.

Since the arguments we follow for \dto{} and \hto{} are similar,
we focus our presentation on \dto{} and leave a detailed exposition
of the results for \hto{} to the \suppinfo{} \cite{supp}.

\ssection{Pseudo-spin model} 
Our ultimate goal is to understand the interactions acting between the
low energy states of the CEF manifold. In \dto{} this is a
Kramers doublet built from states in the $\dgs$ manifold of the $4f^9$
configuration of \dth{}.  For a realistic CEF potential
\cite{rosenkranz2000crystal,bertin2012crystal}, the ground state is a
doublet with a gap of $\sim 30\meV$ to the first excited level. This
doublet transforms in the $\Gamma_5\oplus \Gamma_6$ representation of
the site symmetry group $D_{3d}$ and has Ising-like character
\cite{huang2014quantum}, as shown below. One writes the
doublet states as
\begin{equation}
  \label{doublet}
  \ket{\pm} = 
  \alpha \ket{\pm \tfrac{15}{2}} 
  \pm \delta_1 \ket{\pm \tfrac{9}{2}}
  -\delta_2 \ket{\pm \tfrac{3}{2}}
  \mp \delta_3 \ket{\mp \tfrac{3}{2}}
  + \delta_4 \ket{\mp \tfrac{9}{2}}
  \pm \delta_5 \ket{\mp \tfrac{15}{2}},
\end{equation}
where $\ket{M} \equiv \ket{{15}/{2},M}$ are are eigenstates of the
$\mJ^2$ and $\mJ^z$ operators of the $\dgs$ manifold. We choose the
doublet basis so that $\delta_2\delta_3-3\delta_1\delta_4 +5\alpha
\delta_5=0$ to make the $\mJ^z$ dipole operator diagonal. The matrix
elements of the magnetic dipole operators then take the form
\begin{align}
  \bra{\sigma} \mJ^{\pm} \ketp{\sigma'} &= 0, & 
  \bra{\sigma} \mJ^z \ketp{\sigma'} &= \lambda \sigma \delta_{\sigma\sigma'},
\end{align}
where $\sigma = \pm$ and the $\lambda$-factor is 
\begin{equation}
\lambda \equiv \frac{15}{2} - 3 \left(
\delta_1^2+2\delta_2^2
+3\delta_3^2+4\delta_4^2+5\delta_5^2\right).
\end{equation}
Including the \dth{} Land\'e factor $g=4/3$, the magnetic moment $\mu
= g \muB \lambda$ is found experimentally to be very close to maximal.
Estimates from the saturation magnetization and high-temperature
susceptibility \cite{flood1974magnetization,fukazawa2002magnetic} give
$\mu = (10 \pm 0.1)\ \muB$. This constrains the $\ket{\pm}$ states in
Eq. (\ref{doublet}) to be predominantly $\ket{\pm 15/2}$
\cite{Ramirez_Nature,rosenkranz2000crystal}. We thus have an upper
bound of $\delta^2_n \lesssim 0.025/n$ given $\pm 0.1\ \muB$ error
bars.

Since the CEF energy scale is large, it is sufficient to first
consider only a bare projection of the microscopic \dth{}-\dth{}
interactions acting within the full $\dgs$ manifold into the ground
doublet.  Such a model can be expressed in terms of the components of a
pseudo-spin $1/2$ operator
\begin{align}
  \mS^{\pm} &= \ket{\pm}\bra{\mp}, & 
  \mS^z    &= \frac{1}{2}\left(\ket{+}\bra{+}-\ket{-}\bra{-}\right).
\end{align}
The symmetry of the pyrochlore lattice constrains the form of any such
projected Hamiltonian. The nearest-neighbor \EE{} model for \dto{}
is restricted to the form \cite{huang2014quantum}
\begin{equation}
\label{xyzxz-model}
  \sum_{\avg{ij}} \left[ 
\exJ_{xx} \mS^x_i \mS^x_j +
 \exJ_{yy} \mS^y_i \mS^y_j +
\exJ_{zz} \mS^z_i \mS^z_j + \exJ_{xz} \left(
\mS^z_i \mS^x_j+\mS^x_i \mS^z_j
\right)
\right].
\end{equation}
Since the $\exJ_{\mu\nu}$ are \emph{not} bond dependent, the
$\exJ_{xz}$ coupling can be removed by a global rotation of the
pseudo-spin operators about the $\hat{y}$ axis
\cite{huang2014quantum}, yielding
\begin{equation}
  \label{xyz-model}
  \sum_{\avg{ij}} \left[ 
\exJ_x \mS^x_i \mS^x_j +
 \exJ_y \mS^y_i \mS^y_j +
\exJ_z \mS^z_i \mS^z_j \right].
\end{equation} 
We will find below that the net $\exJ_{zz}$, including both the
contribution from microscopic inter-ionic super-exchange and the
nearest-neighbor MDD contribution, is the dominant coupling and that
it is positive, as originally found empirically in
Ref. [\onlinecite{denhertog}].  At leading order, the rotated
couplings are
\begin{align}
    \label{rotations}
  \exJ_x &\sim \exJ_{xx} - \frac{\exJ_{xz}^2}{\exJ_{zz}}, &
  \exJ_y &= \exJ_{yy}, &
  \exJ_z &\sim \exJ_{zz} + \frac{\exJ_{xz}^2}{\exJ_{zz}}. &
\end{align}
This rotation affects the relationship between the components of the
dipole moment and those of the pseudo-spin $1/2$ at
$O(\exJ_{xz}/\exJ_{zz})$.  We characterize deviations from the Ising
limit using the nearest-neighbor transverse scale $\exJtr \equiv
(\exJ_x+\exJ_y)/4$ \cite{huang2014quantum}.  Our goal is to estimate
the size of $\exJtr$ and thus the strength of the quantum effects within the
low-energy pseudo-spin 1/2 model.

\ssection{Interactions in rare-earth pyrochlores} To understand how
the microscopic inter-ionic coupling mechanisms generate $\exJtr$, we
need to consider higher energy physics and a description beyond the
lowest CEF doublet.  At the atomic level, there is a hierarchy of
energy scales for the rare-earth ions: the Coulomb interaction being
largest, followed by spin-orbit coupling and the CEF potential. The
Coulomb interaction plays two roles: it provides the repulsion that
separates the different $4f^n$ charge states and splits each $4f^n$
manifold in approximate fixed orbital and spin angular momenta
$\termLS$ \cite{condon1951theory}.  Spin-orbit coupling then lifts the
degeneracy of the states with the same total angular momentum $\rm J$,
giving the terms $\termLSJ$. Finally, the single-ion spherical
symmetry of the ion is broken by the CEF and the $\termLSJ$ manifolds
are further split into CEF multiplets.

We thus move one abstraction level up from the pure ground 
doublet, considering interactions within the $\dgs$ manifold. Due to
the spherical symmetry of the free ions, it is useful to categorize
interactions and operators in terms of their \emph{rank} as spherical
tensors. Due to the large value $\tamJ=15/2$ in the $\dgs$ manifold of
\dto{}, multipolar interactions up to rank-15 are possible
\cite{santini2009multipolar}.  Interactions at this level will
typically be on the order of $\sim 1\K$, as inferred from the
Curie-Weiss scale \cite{Ramirez_Nature,Harris_PRL}.  Since the CEF energy scale is of order
$300\K$ \cite{rosenkranz2000crystal}, it is sufficient to consider
only the projection into the ground doublet. Perturbative corrections
are then expected at second order \cite{molavian2007dynamically} with
scale of $\sim (1\K)^2/300\K \sim 3\mK-5\mK$ and thus can be
neglected.  There are several sources of interactions between
rare-earth ions: electro- and magnetostatic interactions,
super-exchange, direct exchange, lattice-mediated exchange, etc.  Of these the MDD and
super-exchange interactions are found to be the most significant
\cite{supp}.

The long-range MDD interaction features prominently in classical spin
ices \cite{siddharthan1999ising,bramwell2001spin,denhertog,Ruff_DTO}. When
projected into the ground doublet, it takes the form
\begin{equation}
  \label{dipolar}
  \exDp \sum_{i<j} \mS^z_i \mS^z_j 
  \left(\frac{r_{nn}}{r_{ij}}\right)^3
  \left[
    \hat{z}_i\cdot \hat{z}_j -
    \left(\hat{r}_{ij} \cdot \hat{z}_i\right)
    \left(\hat{r}_{ij} \cdot \hat{z}_j\right)
    \right],
\end{equation}
where $r_{nn}$ is the nearest-neighbor distance between the rare-earth
atoms, $\vec{r}_{ij} \equiv \vec{r}_i-\vec{r}_j$ is the displacement
vector between the two sites and $\hat{z}_i$ is the quantization axis
along the local cubic $[111]$ direction at site $\vec{r}_i$.  The
strength of the coupling is characterized by $\exDp = g^2 \lambda^2
\muB^2 \mu_0/(4\pi r_{nn}^3)$. The largest piece is the
nearest-neighbor contribution \cite{Yavorski_DTO}, with coefficient
$\exD \equiv 5\exDp/3$ which is $\sim 8.9\K$ in \dto{} \footnote{ Note
that some prior work uses a different convention for their Ising variables,
taking values $\pm 1$ rather than the pseudo-spin $\pm 1/2$ used here.
This accounts for a factor of four difference in the stated exchange
constants $\exD$ and $\exJ$.}.  To obtain a quantitative agreement
with experiments, corrections to the Ising couplings must be
included. The nearest-neighbour correction $J$ is the largest,
estimated to be $\sim 4.96\K$ in \dto{} \cite{denhertog}, later
refined to $\sim 4.55\K$ \cite{Yavorski_DTO}. This coupling has usually
been attributed to super-exchange processes between the rare-earth
ions and provides an indication of their overall scale \footnote{The
sign convention for $\exJ$ is reversed relative to some prior work in
Refs.~\cite{denhertog,melko2001long,Ruff_DTO,Melko_JPCM}.}.  We are thus confronted
with the question: with such a large scale,
why does super-exchange not generate significant
transverse couplings $\exJtr$?

\ssection{Oxygen-mediated super-exchange}
To answer this question, we consider super-exchange interactions
proceeding through the oxygen atoms that surround each rare-earth
ion. To simplify our notation, we single out one exchange path
consisting of two rare-earths and one oxygen. For the rare-earth
sites, we define the creation operators $\h{f}_{1\alpha}$ and
$\h{f}_{2\alpha}$ while for the intermediate oxygen site we use
$\h{p}_{\alpha}$.  A combined spin-orbital index $\alpha \equiv
(m,\sigma)$ where $m$ and $\sigma$ label the orbital and spin quantum
numbers is used throughout. Due to the localized nature of the
rare-earth ions, we start with the atomic Hamiltonian at each site
\begin{equation}
  H_0 \equiv H_{f,1}+H_{f,2} + H_p,
\end{equation}
where $H_{f,1}$ and $H_{f,2}$ are the atomic Hamiltonian for the two
rare-earth ions and $H_p$ for the oxygen site. On the oxygen site, we
consider only the atomic potential $\DeltaOx$ and the cost to place
two holes together on the same oxygen, $U_p$. We will not invoke the
details of the rare-earth atomic Hamiltonian, aside from selecting the
$\dgs$ manifold of the $4f^9$ configuration.  We perturb $H_0$ with
the hybridization terms
\begin{equation}
V \equiv \sum_{\alpha\beta} \left[
 {t}_{1,\alpha\beta} \h{f}_{1\alpha} p^{}_{\beta} +
 {t}_{2,\alpha\beta} \h{f}_{2\alpha} p^{}_{\beta} +{\rm h.c.}
\right],
\end{equation}
that represent electron hopping between the orbitals of the rare-earth and
oxygen ions. Super-exchange interactions are generated at
fourth-order in perturbation theory in $V$ \cite{onoda2011quantum}.
For the details of the charge-transfer processes, see
the \suppinfo{} \cite{supp}.

To simplify the resolvents that appear in the perturbative expansion
\cite{lindgren1974rayleigh,supp}, we follow Ref. [\onlinecite{onoda2011quantum}] and make
the common approximation that only the charging energies $E(f^{9 \pm
1}) - E(f^9) \equiv U^{\pm}_f$ are significant. This neglects all
other intra-atomic splittings on the rare-earth sites, such as those
due to Hund's coupling or spin-orbit coupling. The energies
$U^{\pm}_f$ are expected to be of order $5-10\eV$ in $4f$ compounds
\cite{van1988electron}.  Within this approximation, the effective
Hamiltonian can be written
\cite{onoda2011quantum}
\begin{equation}
H_{\rm eff} = 
 \sum_{\alpha\beta\mu\nu}
\left(P_{1} \h{f}_{1\alpha} f^{}_{1\beta} P_{1}\right) 
\mathcal{I}^{\alpha\beta\mu\nu}_{12}
\left(P_{2} \h{f}_{2\mu} f^{}_{2\nu}  P_{2}\right),
\end{equation}
where $P_{i}$ projects into the ground state manifold of $H_{f,i}$ at
site $i$. The interaction matrix $\mathcal{I}$ is defined as
\cite{supp}
\begin{eqnarray}
\label{super-exchange-model}
&&\mathcal{I}^{\alpha\beta\mu\nu}_{12} \equiv
\frac{2}{(U^+_f +\DeltaOx)^2} 
\Bigg[
- \frac{2
\left[
{t}^{}_{2}
\h{t}_{2}
\right]^{\mu\nu}
\left[
{t}^{}_{1}
\h{t}_{1}\right]^{\alpha\beta}
}{2U^+_f +U_p +2\DeltaOx} +
\nonumber \\
&& 
\left(\frac{1}{U^+_f+U^-_f}+\frac{2}{2U^+_f +U_p +2\DeltaOx}\right)
\left[{t}^{}_{2} \h{t}_{1} \right]^{\mu\beta}
\left[{t}^{}_{1} \h{t}_{2} \right]^{\alpha \nu}
\Bigg].
\end{eqnarray}
Here we dropped constants and single site terms that serve only to
renormalize the on-site single-ion Hamiltonian.

Generically, the interactions $\mathcal{I}$ are complicated due to the
$f^9$ states that make up the $\dgs$ ground state manifold of
\dth{}. However, a strong constraint arises from the one-electron form
$\h{f}_{\alpha} f_{\beta}$ taken by the operators at each rare-earth
site. When projected into the $\dgs$ manifold, these operators can
generate only a small subset of all possible interactions between the
$\tamJ=15/2$ degrees of freedom of \dth{}.  This can be seen
explicitly by considering how these operators transform under
rotations. Since each $f$ electron carries a total angular momentum of
$\tamJ=5/2$ or $\tamJ=7/2$, the operators $\h{f}_{\alpha} f_{\beta}$
are multipoles with rank ranging from rank-0 ($5/2-5/2$ or $7/2-7/2$)
up to rank-7 ($7/2+7/2$).  Hence, we find that \emph{only
multipolar interactions up to rank-7 are generated by super-exchange}.  
Because of the spherical symmetry of the free ions; the
complicated intra-atomic interactions are rank-0 operators and thus
\emph{do not} change the rank counting above \cite{supp}.

Including the CEF splittings in the resolvents only changes this
result slightly. Given the small overall splitting $\DeltaCF \lesssim
100 \meV$ induced by the CEF
\cite{rosenkranz2000crystal,bertin2012crystal}, we can treat this as a
perturbation to $H_0$ along with $V$. Since the CEF potential is a
one-electron operator and is time-reversal invariant, it contains only
operators up to rank-6 for the same reasons as discussed above
\cite{stevens1952matrix}. Each inclusion of the CEF operator can thus
increase the rank by 6, with up to rank-13 operators at order $\sim
\DeltaCF/U^{\pm}_f$ , and operators up to the maximal rank-15 at order
$\sim (\DeltaCF/U^{\pm}_f)^2$. The rank-15 operators that directly
link the leading $\ket{\pm 15/2}$ components are thus \emph{strongly
suppressed} by a factor of $(\DeltaCF/U^{\pm}_f)^2 \sim 10^{-4}$.  We
can thus ignore these small corrections and can consider only
interactions of rank-7 or lower.

\ssection{General argument}
Returning to a more general physical picture, the above results allow
us to explain why quantum effects are small in the classical spin
ices \dto{} and \hto{}. Given the dominant $\ket{\pm 15/2}$ or $\ket{\pm 8}$ 
CEF ground states in \dth{} and \hth{} \cite{rosenkranz2000crystal},
a large transverse scale $\exJtr$ must originate from \emph{very}
high-rank multipole interactions. These multipole interactions are
only generated weakly from super-exchange contributions, as they are
much higher than rank-7. Due to this suppression, the leading
contributions to the transverse couplings must come from the
sub-dominant spectral components of the CEF ground doublet, the
$\delta_n$ components given in Eq. (\ref{doublet}) for \dto{}. When
these are included, generation of $\exJtr$ by super-exchange becomes
feasible. As a rough estimate, if $J$ is the typical scale of
contributions from super-exchange then we expect
\begin{align} 
  \label{scaling}
  \exJtr &\sim  \left(\frac{\delta}{\alpha}\right)^2 \exJ, & \exJ_{z} \sim -\exJ,
\end{align}
where $\delta$ represents the maximum size of the sub-dominant
components of the CEF doublet.  Each transverse pseudo-spin operator
$S^{\pm}$ contributes one factor of $\delta/\alpha$ to the scale
$\exJtr$ \footnote{ For the case of the $\Gamma_5\oplus \Gamma_6$
doublet of \dto{}, the $\exJ_{xz}$ exchange will be of order
$\delta/\alpha$, since only one of the pseudo-spins is
transverse. Redefining the axes by the rotation (\ref{rotations}), we
recover the estimate as stated.}. Since we can constrain
$\delta/\alpha \lesssim 0.1$ via the size of the magnetic moment, then
one expects, all things being equal, the transverse
scale $\exJtr$ to be suppressed by two orders of magnitude relative to
the super-exchange contributions to the Ising coupling $\exJ_z$. Since
we know $\exJ \sim 5\K$ in \dto{} \cite{denhertog,Yavorski_DTO}
and $\exJ \sim 2\K$ in \hto{} \cite{bramwell2001spin}, this implies
$\exJtr \lesssim 50\mK$.  This is the main conclusion of
our work.

\ssection{Estimate from model calculation}
The simple scaling argument above does not include combinatoric
factors or other unforeseen quirks or cancellations that could favor
or disfavor the generation of transverse $\exJtr$ couplings. To check
that such anomalies do not occur, we carry out an explicit calculation
to verify this scaling estimate. We work within the charging
approximation as encapsulated in Eq. (\ref{super-exchange-model}),
considering only a single super-exchange path.

The shortest path between the rare-earth ions and oxygen passes
through the $O'$ site situated in the center of each tetrahedron.
Within the Slater-Koster approximation \cite{slater1954simplified},
the hopping matrices $t_{1}$, $t_2$ can be expressed as $ t_{i} =
\h{R}_{i} t_0 $ where $R_{i}$ is a rotation of the $f$ and $p$
orbitals that takes a set of axes aligned to the $i$ local axes into
the global frame.  The matrices $t_0$ define overlaps in the frame
aligned along the bond axis and take the simple form $[t_0]_{mm'} =
\delta_{mm'} (\delta_{|m|=1} t_{pf\pi} + \delta_{m=0}
t_{pf\sigma})$. We note that $t_{pf\pi}$ is smaller in magnitude than
$t_{pf\sigma}$ and of opposite sign \cite{takegahara1980slater},
though their precise values will not be important to our discussion.
Since $t_0$ is diagonal and $R_i$ is unitary, the products that appear
in Eq. (\ref{super-exchange-model}) are
\begin{align}
  \label{effective-hopping}
  \left[t^{}_i \h{t}_j\right]^{\alpha\beta} &= 
t^{\alpha}_0 t^{\beta}_0 \left[\h{R}_i R^{}_j\right]^{\alpha\beta}, &
  \left[t^{}_i \h{t}_i\right]^{\alpha\beta} &= 
\left(t^{\alpha}_0 \right)^2 \delta_{\alpha\beta},
\end{align}
as in Ref. [\onlinecite{onoda2011quantum}].  Consideration of further
super-exchange paths between the other oxygen atoms would change the
form of $t_{i}$, but not the overall structure of the interactions.

To evaluate the super-exchange interactions, we proceed as follows: we
first construct the CEF ground doublet states of Eq. (\ref{doublet})
from the full $f^9$ manifold. Next, we project the one-electron
operators $\h{f}_{\alpha} f_{\beta}$ into this basis. Finally we sum
these one-electron operators with the interaction constants of
Eq. (\ref{super-exchange-model}) evaluated using the $t_i$ given above
for single bond of the lattice. The remaining bonds can be recovered
using the lattice symmetry.  This gives a model of the form shown in
Eq. (\ref{xyzxz-model}). At leading order in the $\delta_n$
coefficients, one has
\begin{align}
\label{se-result}
 \exJ_{xx} &=-4 \exJ \left(\frac{\delta_5}{\alpha}\right)^2 + O(\alpha \delta^3), &
  \exJ_{xz} &= +2 \exJ \left(\frac{\delta_5}{\alpha}\right)
           + O(\alpha \delta^2),\nonumber \\ 
  \exJ_{zz} &= -\exJ 
           + O(\alpha \delta^2), &
  \exJ_{yy} &= 0+O(\alpha \delta^3), 
\end{align}
where all of the $\delta_n$ have been considered at the same order in
the expansion \footnote{ To perform this expansion we treat all
$\delta_n$ as equally small. Practically speaking, we replace
$\delta_n \rightarrow \eta \delta_n$ and expand to leading order in
$\eta$, then send $\eta \rightarrow 1$. Statements such as
$O(\delta^2)$ imply the dropped terms involve a product of some pair
of the $\delta_n$.}.  The super-exchange energy scale $\exJ$ is
\begin{align}
\nonumber
\exJ =
&\frac{2\alpha^4 t^4_{pf\sigma}}{(U^+_f+\Delta)^2}
\left(\frac{1}{U^+_f+U^-_f}+\frac{2}{2U^+_f + U_p +2\Delta}\right) \times
\\
& \frac{1}{2187}\left(
121+96\left(\frac{t_{pf\pi}}{t_{pf\sigma}}\right)^2+
450\left(\frac{t_{pf\pi}}{t_{pf\sigma}}\right)^4\right).
\label{se-scale}
\end{align}
From Eqs. (\ref{se-result}) and (\ref{se-scale}), we see that our
na\"ive scaling argument of $\exJtr \sim (\delta/\alpha)^2 \exJ$ does
in fact hold in this more detailed calculation.  We also note that the
sign of the super-exchange contribution to $\exJ_{zz}$ is negative, as
required for \dto{} \cite{denhertog,Yavorski_DTO}.

Only the super-exchange contribution to $\exJ_{zz}$ has been computed
here; when rotating to eliminate the $\exJ_{xz}$ coupling, we must
also include the large contribution $D$ to $\exJ_{zz}$ coming from the
nearest-neighbor part of the MDD interactions.  We thus shift
$\exJ_{zz} \rightarrow \exJ_{zz} +\exD$ where $\exD$ is given in
Eq. (\ref{dipolar}).  Using the approximate rotations of
Eq. (\ref{rotations}), we find $\exJ_y \sim 0$ and
\begin{align}
  \exJ_x &= -\frac{4 {\exD} \exJ}{{\exD}-\exJ}\left(\frac{\delta_5}{\alpha}\right)^2, &
  \exJ_z & = \exD-\exJ + \frac{4 \exJ^2}{\exD-\exJ}\left(\frac{\delta_5}{\alpha}\right)^2.
\end{align}
From the constraint on the moment size, we
find $\delta^2_5/\alpha^2 \leq 0.005$ so then $\exJtr \lesssim
56\mK$. At the temperature scale of interest, monopole excitations
\cite{Castelnovo_Nature} are suppressed and quantum effects can
proceed only through quantum tunnelling \emph{within} the spin ice
manifold \cite{hermele2004pyrochlore}.  The strength of tunnelling $g$
appears perturbatively at third-order in the transverse coupling as $g
\sim 12 \exJtr^3/\exJ_z^2$
\cite{hermele2004pyrochlore,huang2014quantum,kato2014numerical}. Using
the above estimate, these quantum effects would only become relevant
for temperatures $\lesssim 0.14\ {\rm mK}$ in \dto{}.  In fact, this
estimate could be reduced significantly depending on how close the
sub-leading spectral components of $\ket{\pm}$ are to their maximal
values allowed by the moment size.  For example, using the CEF
parameters of \citet{bertin2012crystal}, one has $\delta_5 \sim
10^{-3}$ which is much smaller than the bound implied by $\delta_n^2
\lesssim 0.025/n$ from the moment, giving a significantly smaller
transverse coupling $\exJtr$. Following the same methods, the
estimate for tunnelling in \hto{} is also $\lesssim 0.1 \mK$
\cite{supp}.

\ssection{Outlook}
We have presented a general argument as to why transverse exchanges in
the canonical spin ices are small. In particular, we have shown that
the generation of the required high rank multipolar couplings needed
to link the nearly maximal $\ket{\pm \tamJ}$ states of the CEF
doublets are strongly suppressed.  Using an approximate treatment of
the oxygen mediated super-exchange interaction, we have provided some
heuristic bounds on this size of these couplings.  The presence of all
these factors can provide an explanation of spin ice behavior in the
related germanates \abo{Ho}{Ge} \cite{hallas2012statics,DHGO} and \abo{Dy}{Ge} \cite{ke2008magnetothermodynamics,DHGO} or the stannates
\abo{Ho}{Sn} \cite{matsuhira2002low,HSO} and \abo{Dy}{Sn} \cite{matsuhira2002low,DSO}.  
These criteria also suggest what
features to look for to move away from the classical spin ice limit
\cite{Gingras_RPP}.  For example, compounds with Kramers ions and
large Ising-like moments, but significant sub-leading components in
their ground state CEF doublet could present quantum behavior at more
experimentally relevant temperature scales \cite{Gingras_RPP}. In this context, the
spinel CdEr$_2$Se$_4$ \cite{lago2010cder} with Ising Er$^{3+}$ moment
is an intriguing example where this logic may apply.

\ssection{Acknowledgements}
We thank Zhihao Hao for useful discussions.  This work was supported
by the NSERC of Canada, the Canada Research Chair program
(M. J. P. G., Tier 1), the Canadian Foundation for Advanced Research
and the Perimeter Institute (PI) for Theoretical Physics.  Research at
PI is supported by the Government of Canada through Industry Canada
and by the Province of Ontario through the Ministry of Economic
Development \& Innovation. M.G. acknowledges the hospitality and
generous support of the Quantum Matter Institute at the University of
British Columbia and TRIUMF where part of this work was completed.

\ssection{Note added} 
After completion of this work, a preprint appeared,
Ref. \cite{iwahara2015exchange}, that reaches some similar conclusions
on the interactions between $\termLSJ$ multiplets in rare-earth ions.

\bibliography{draft}

\newpage


\section*{Supplementary material}

# Supplemental Material for "How quantum are classical spin ices?"

Jeffrey G. Rau[1] and Michel J. P. Gingras[1,2,3]

[1]*Department of Physics and Astronomy,*
*University of Waterloo, Ontario, N2L 3G1, Canada*

[2]*Perimeter Institute for Theoretical Physics,*
*Waterloo, Ontario, N2L 2Y5, Canada*

[3]*Canadian Institute for Advanced Research, 180 Dundas Street West,*
*Suite 1400, Toronto, ON, M5G 1Z8, Canada*

# TRANSVERSE INTERACTIONS IN HOLMIUM TITANATE

## Ground state doublet

In the case of $Ho^{3+}$, the crystal field (CEF) ground state is a non-Kramers doublet built from the $^5I_8$ manifold of the $4f^{10}$ configuration [1, 2]. The first excited CEF level is $\sim 20$ meV above the ground doublet [2]. This non-Kramers doublet transforms in the $E_g$ representation of the $D_{3d}$ site symmetry group and can be written

$$|\pm\rangle = \alpha\,|\pm 8\rangle \mp \delta_1\,|\pm 5\rangle + \delta_2\,|\pm 2\rangle \mp \delta_3\,|\mp 1\rangle + \delta_4\,|\mp 4\rangle \mp \delta_5\,|\mp 7\rangle \tag{1}$$

where $|M\rangle \equiv |8, M\rangle$ are eigenstates of the $J^2$ and $J^z$ operators of the $^5I_8$ manifold. This doublet is naturally Ising-like with

$$\langle\sigma|J^\pm|\sigma'\rangle = 0, \qquad\qquad \langle\sigma|J^z|\sigma'\rangle = \lambda\sigma\delta_{\sigma\sigma'}, \tag{2}$$

where $\sigma = \pm$ and the $\lambda$-factor is

$$\lambda \equiv 8 - 3\left(\delta_1^2 + 2\delta_2^2 + 3\delta_3^2 + 4\delta_4^2 + 5\delta_5^2\right). \tag{3}$$

Including the $Ho^{3+}$ Landé factor, $g = 5/4$, the magnetic moment $\mu = g\mu_B\lambda$ is found to be very close to the maximal $10\mu_B$ experimentally [1, 3, 4]. This constrains the spectral content of the $|\pm\rangle$ states in Eq. (1) to be predominantly $|\pm 8\rangle$. We can also appeal to more detailed information on the CEF structure. Fits to the CEF levels obtained via neutron scattering have yielded estimates of the sub-leading components $\delta_n$. For example, work by Bertin *et al.* [5] gives

$$\delta_1 = 0.189, \qquad \delta_2 = 0.014, \qquad \delta_3 = 0.070, \qquad \delta_4 = 0.031, \qquad \delta_5 = 0.005. \tag{4}$$

Rather than rely on these parameters, we shall consider the bound implied by a deviation of the moment from maximal by some amount $\epsilon$. In $Dy_2Ti_2O_7$, we used $\epsilon = 0.01$ due to the $\pm 0.1\mu_B$ error bar from Ref. [6]. For $Ho_2Ti_2O_7$ this relation is $\delta_n^2 \lesssim 8\epsilon/(3n)$, but given the lack of stated bound on the moment, we use a more conservative $\epsilon = 0.05$ which implies $\delta_n^2 \lesssim 0.133/n$.

## Effective pseudo-spin 1/2 model

We proceed similarly to the $Dy_2Ti_2O_7$ case in the main text. Since the CEF energy scale is large, it is sufficient to first consider only a bare projection of the microscopic $Ho^{3+}$-$Ho^{3+}$ interactions

2

acting within the full $^5I_8$ manifold into the ground doublets. Such a model can be expressed in terms of the pseudo-spin 1/2 operators

$$S^{\pm} = |\pm\rangle\langle\mp|, \qquad S^z = \frac{1}{2}(|+\rangle\langle+| - |-\rangle\langle-|). \tag{5}$$

acting within the low-energy Hilbert space spanned by the $|\pm\rangle$ doublets. The symmetry of pyrochlore lattice constrains the form of any such projected Hamiltonian. For a non-Kramers doublet, a nearest-neighbor model can be shown to be restricted to the form [7, 8]

$$\sum_{\langle ij \rangle} \left[ J_{zz} S_i^z S_j^z - J_{\pm}\left(S_i^+ S_j^- + S_i^- S_j^+\right) + J_{\pm\pm}\left(\gamma_{ij} S_i^+ S_j^+ + \gamma_{ij}^* S_i^- S_j^-\right) \right]. \tag{6}$$

where the bond dependent phases $\gamma_{ij}$ are defined in Ref. [8]. Once dipolar interaction are included, we will find that $J_{zz}$ is the dominant bare coupling and that it is positive. We will characterize deviations from the Ising limit using the nearest-neighbor transverse scale $J_{\pm}$, as quantum effects from $J_{\pm\pm}$ appear at higher order in peturbation theory [8, 9]. Our goal is to estimate the size of $J_{\pm}$ which is responsible for the quantum dynamics within the low-energy pseudo-spin 1/2 model.

**Ising interactions**

As in $Dy_2Ti_2O_7$, the long-range dipole-dipole interactions play an important role, In $Ho_2Ti_2O_7$ these take the same form

$$\mathcal{D} \sum_{ij} S_i^z S_j^z \left(\frac{r_{nn}}{r_{ij}}\right)^3 \left[\hat{z}_i \cdot \hat{z}_j - \left(\hat{r}_{ij} \cdot \hat{z}_i\right)\left(\hat{r}_{ij} \cdot \hat{z}_j\right)\right], \tag{7}$$

where $r_{nn}$ is the nearest-neighbor distance between the rare-earth atoms, $\vec{r}_{ij} \equiv \vec{r}_i - \vec{r}_j$ is the displacement vector between the two sites and $\hat{z}_i$ is the quantization axis along the local cubic [111] direction at site $\vec{r}_i$. The strength of the coupling is characterized by $\mathcal{D} = g^2 \lambda^2 \mu_B^2 \mu_0/(4\pi r_{nn}^3)$. The largest piece is the nearest-neighbor contribution, with coefficient $D \equiv 5\mathcal{D}/3$ which is $\sim 9.4$ K in $Ho_2Ti_2O_7$ [10], slightly larger than in $Dy_2Ti_2O_7$. As in $Dy_2Ti_2O_7$, to obtain a quantitative agreement with experiments, corrections to the Ising couplings must be included. The nearest-neighbor correction $J$ is the largest, estimated to be $\sim 2.08$ K [11] for $Ho_2Ti_2O_7$.

**Model calculation**

We follow the same procedure as in the main text for $Dy_2Ti_2O_7$, this time working with the states defined in Eq. (1) drawn from the $4f^{10}$ manifold appropriate for $Ho^{3+}$. Aside from the

3

construction of these states, the methods are identical. We find

$$J_{\pm} = +J \cdot \rho_{\pm} \left(\frac{t_{pf\pi}}{t_{pf\sigma}}\right)\left(\frac{\delta_5}{\alpha}\right)^2 + O(\alpha\delta^3), \tag{8a}$$

$$J_{\pm\pm} = -J \cdot \rho_{\pm\pm} \left(\frac{t_{pf\pi}}{t_{pf\sigma}}\right)\left(\frac{\delta_5}{\alpha}\right)^2 + O(\alpha\delta^3), \tag{8b}$$

$$J_{zz} = -J + O(\alpha^3\delta), \tag{8c}$$

where all of the $\delta_n$ have been considered at the same order in the expansion. The super-exchange energy scale $J$ is given by

$$J = \frac{4\alpha^4 t_{pf\sigma}^4}{(U_f^+ + \Delta_{pf})^2} \left(\frac{1}{U_f^+ + U_f^-} + \frac{2}{2U_f^+ + U_p + 2\Delta_{pf}}\right) \frac{1}{2187} \left(121 + 48\left(\frac{t_{pf\pi}}{t_{pf\sigma}}\right)^2 + 207\left(\frac{t_{pf\pi}}{t_{pf\sigma}}\right)^4\right). \tag{9}$$

The $\rho_{\pm}$ and $\rho_{\pm\pm}$ are rational functions of the ratio $x \equiv t_{pf\pi}/t_{pf\sigma}$ defined as

$$\rho_{\pm}(x) = \frac{9x^2(160 - 8x + 7x^2)}{8(121 + 48x^2 + 207x^4)}, \qquad \rho_{\pm\pm}(x) = \frac{9x^2(68 - 16x - 7x^2)}{16(121 + 48x^2 + 207x^4)}. \tag{10}$$

Reasonable values of the Slater-Koster ratio satisfy $-1 < t_{pf\pi}/t_{pf\sigma} < 0$. For this range, we can bound these functions as

$$0 < \rho_{\pm}(t_{pf\pi}/t_{pf\sigma}) \lesssim 0.6, \qquad 0 < \rho_{\pm\pm}(t_{pf\pi}/t_{pf\sigma}) \lesssim 0.15. \tag{11}$$

As in $Dy_2Ti_2O_7$, we see that scaling from the main text, $J_{\pm} \sim (\delta/\alpha)^2 J$ holds with no large combinatoric factors or spurious cancellations. We also note that the sign of the super-exchange contribution to the $J_{zz}$ is again, correctly negative [10]. Setting the super-exchange scale to $J = 2.08$ K from the experimental constraints [10], we then can estimate $J_{\pm}$ and $J_{\pm\pm}$ as

$$J_{\pm} \lesssim 1.25 \text{ K} \cdot \left(\frac{\delta_5}{\alpha}\right)^2, \qquad |J_{\pm\pm}| \lesssim 0.31 \text{ K} \cdot \left(\frac{\delta_5}{\alpha}\right)^2. \tag{12}$$

For a moment size that deviates from maximal by $\epsilon = 0.05$, this ratio is bounded as $(\delta_5/\alpha)^2 \lesssim 0.027$ giving

$$J_{\pm} \lesssim 34 \text{ mK}, \qquad |J_{\pm\pm}| \lesssim 8.5 \text{ mK}. \tag{13}$$

Here the relevant transverse scale is the $J_{\pm}$ that appears in the nearest-neighbor model (6). With $J_z = D - J \sim 7.32$ K, the implied tunnelling scale $g$ is $\ll 0.1$ mK and thus is not experimentally relevant. We note that these estimates for $J_{\pm}$ and $J_{\pm\pm}$ scale linearly in the deviation from the maximal moment. For the CEF parameters of Ref. [5] we have $(\delta_5/\alpha)^2 \sim 10^{-6}$, and one would have to look at next order in the $\delta_n$ to get at the true leading behavior. It is unclear if the fitting of the CEF energy levels is sufficiently precise to provide a useful estimate of the true size of $\delta_5$.

**GENERAL DISCUSSION OF SUPER-EXCHANGE**

We use the notation of the main text. As discussed there, we consider the bare Hamiltonian of the two rare-earth sites and the intermediate oxygen atom

$$H_0 \equiv H_{f,1} + H_{f,2} + H_p. \qquad (14)$$

We perturb this with the hybridization terms

$$V \equiv \sum_{\alpha\beta} \left[ t^\dagger_{1,\alpha\beta} p^\dagger_\alpha f_{1\beta} + t_{1,\alpha\beta} f^\dagger_{1\alpha} p_\beta + t^\dagger_{2,\alpha\beta} p^\dagger_\alpha f_{2\beta} + t_{2,\alpha\beta} f^\dagger_{2\alpha} p_\beta \right], \qquad (15)$$

Exchange interactions between the rare-earth ions come in at fourth-order in $V$. We will follow Lindgren [12] and write the effective Hamiltonian using the ground state projector $P$ and resolvent operator $R$

$$P = \sum_{\alpha \in 0} |\alpha\rangle \langle \alpha|, \qquad R = \sum_{a \neq 0} \frac{|a\rangle \langle a|}{E_0 - E_a}, \qquad (16)$$

where $|\alpha\rangle$ are ground states of $H_0$ with energy $E_0$ and $|a\rangle$ are the excited states of $H_0$ with energies $E_a$. The fourth-order effective Hamiltonian is then [12]

$$H_{\text{eff}} = PVRVRVRVP - PVR^2VPVRVP - PVR^2VRVPVP$$
$$- PVRVR^2VPVP + PVR^2VPVPVP \qquad (17)$$

Now since $PVP = 0$, we can drop all but the two terms $PVRVRVRVP$ and $PVR^2VPVRVP$. The second of these factors as $(PVR^2VP)(PVRVP)$ and serves to cancel non-extensive contributions that arise from the first term.

We thus can simply use $H_{\text{eff}} \sim PVRVRVRVP$, taking to care to consider only this set of (extensive) perturbative processes. At lowest order, we need only consider the excited configurations that can be built by moving at most two electrons between the rare-earth and oxygen sites. These types of intermediate states can be understood by considering the possible values for the oxygen electron number. The range of accessible states are illustrated in Figures 1, 2 and 3. Starting from the ground state $f^n p^6 f^n$ configuration, one follows the colored arrows to trace out a given exchange path. Spin and orbital indices on the $f$ and $p$ electron operators have been suppressed for clarity. We see only two distinct types of processes are needed at fourth order, process 1 (Figure 1) which creates at most one hole on the oxygen and processes 2 and 3 (Figure 2 and Figure 3) which create an intermediate state with two holes on the oxygen. A second set of equivalent paths

5

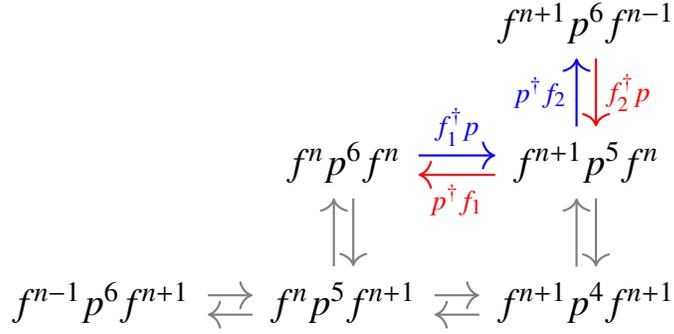

FIG. 1: Process 1 involves intermediate states with only a single hole on the oxygen site. Scales roughly as $\sim t^4(U_f + \Delta_{pf})^{-2}(2U_f)^{-1}$

can be obtained by switching the role of sites 1 and 2 in the processes. Process 1 corresponds to the pieces of $H_{\text{eff}}$ of the form

$$P\left[p^\dagger t_1^\dagger f_1\right] R \left[f_2^\dagger t_2 p\right] R \left[p^\dagger t_2^\dagger f_2\right] R \left[f_1^\dagger t_1 p\right] P. \tag{18}$$

Each appearance of the resolvent operator $R$ can be simplified given the restrictions created by the hybridization operator $V$. For example, the piece $\left[f_1^\dagger t_1 p\right] P$ can only involve states of the form $f^{n+1} p^5 f^n$ since the ground state manifold contains only states of the form $f^n p^6 f^n$. When acting on these states, the resolvent is simplified to an operator that acts only on the states of the rare-earth at site 1

$$R_{f,1} \equiv \sum_{a \in f^{n+1}} \frac{|a\rangle_1 \langle a|_1}{E_{f,a} - E_{f,0} + \Delta_{pf}} \tag{19}$$

where $|a\rangle$ are states of the $f^{n+1}$ configuration and $E_{f,a} - E_{f,0}$ are their energies relative to the $f^n$ ground state, and $\Delta_{pf}$ is the atomic potential of the oxygen site. An identical operator $R_{f,2}$ can be defined that acts only on site 2. The innermost resolvent can be simplified in the same way. Here, both rare-earth sites are involved, but the oxygen state is not. It can be replaced by the operator

$$R_{ff}^{(1)} \equiv \sum_{a \in f^{n+1}} \sum_{a' \in f^{n-1}} \frac{|a\rangle_1 \langle a|_1 |a'\rangle_2 \langle a'|_2}{E_{f,a} + E_{f,a'} - 2E_{f,0}}. \tag{20}$$

This allows us to write the process 1 contribution as

$$P\left[p^\dagger t_1^\dagger f_1\right] R_{f,1} \left[f_2^\dagger t_2 p\right] R_{ff}^{(1)} \left[p^\dagger t_2^\dagger f_2\right] R_{f,1} \left[f_1^\dagger t_1 p\right] P. \tag{21}$$

Now since the resolvents no longer depend on the oxygen degrees of freedom we can regroup this

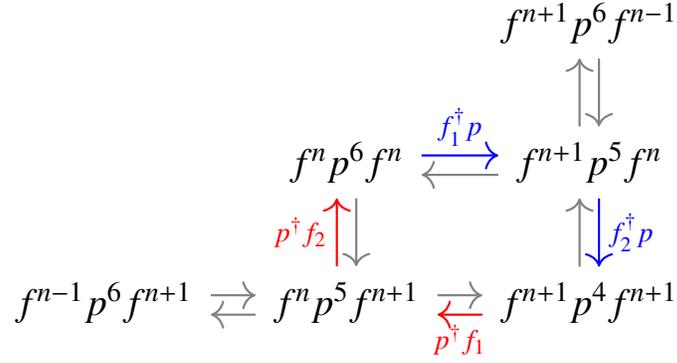

FIG. 2: Process 2 involves intermediate states with two holes on the oxygen site. Scales roughly as $\sim t^4(U_f + \Delta_{pf})^{-2}(2U_f + 2\Delta_{pf} + U_p)^{-1}$

into oxygen and rare-earth parts

$$\sum_{\alpha\beta\mu\nu}\sum_{\alpha'\beta'\mu'\nu'}\left[P_p p_\alpha^\dagger p_\beta p_\mu^\dagger p_\nu P_p\right]\left[P_f f_{1\alpha'} R_{f,1} f_{2\beta'}^\dagger R_{ff}^{(1)} f_{2\mu'} R_{f,1} f_{1\nu'}^\dagger P_f\right]\left[t_{1,\alpha\alpha'}^\dagger t_{2,\beta'\beta} t_{2,\mu\mu'}^\dagger t_{1,\nu'\nu}\right], \quad (22)$$

where we have broken up the projector $P$ into rare-earth and oxygen parts as $P \equiv P_p P_f$. The oxygen part is trivial with $P_p p_\alpha^\dagger p_\beta p_\mu^\dagger p_\nu P_p = \delta_{\mu\nu}\delta_{\alpha\beta}$. The final contribution from process 1 is then given by

$$P_f\left(\sum_{\alpha\beta\mu\nu} f_{1\alpha} R_{f,1} f_{2\beta}^\dagger R_{ff}^{(1)} f_{2\mu} R_{f,1} f_{1\nu}^\dagger \left[t_2 t_1^\dagger\right]^{\beta\alpha}\left[t_1 t_2^\dagger\right]^{\nu\mu}\right)P_f. \quad (23)$$

We carry out the same argument for process 2. Following the arrows in Figure 2 we find the contribution is

$$P\left[p^\dagger t_2^\dagger f_2\right]R\left[p^\dagger t_1^\dagger f_1\right]R\left[f_2^\dagger t_2 p\right]R\left[f_1^\dagger t_1 p\right]P. \quad (24)$$

The outermost resolvents can simplified in the way. The innermost resolvent needs a new definition

$$R_{ff}^{(2)} \equiv \sum_{a\in f^{n+1}}\sum_{a'\in f^{n+1}} \frac{|a\rangle_1 \langle a|_1 |a'\rangle_2 \langle a'|_2}{E_{f,a} + E_{f,a'} - 2E_{f,0} + 2\Delta_{pf} + U_p}, \quad (25)$$

to take into account the additional oxygen holes and Coulomb repulsion in the $f^{n+1}p^4 f^{n+1}$ states, as discussed in the main text. This then gives the operator

$$P\left[p^\dagger t_2^\dagger f_2\right] R_{f,2}\left[p^\dagger t_1^\dagger f_1\right] R_{ff}^{(2)}\left[f_2^\dagger t_2 p\right] R_{f,1}\left[f_1^\dagger t_1 p\right] P. \quad (26)$$

Once again we can factor out the oxygen part as

$$\sum_{\alpha\beta\mu\nu}\sum_{\alpha'\beta'\mu'\nu'}\left[P_p p_\alpha^\dagger p_\beta^\dagger p_\mu p_\nu P_p\right]\left[P_f f_{2\alpha'} R_{f,1} f_{1\beta'} R_{ff}^{(2)} f_{2\mu'}^\dagger R_{f,1} f_{1\nu'}^\dagger P_f\right]\left[t_{2,\alpha\alpha'}^\dagger t_{1,\beta\beta'}^\dagger t_{2,\mu'\mu} t_{1,\nu'\nu}\right]. \quad (27)$$

7

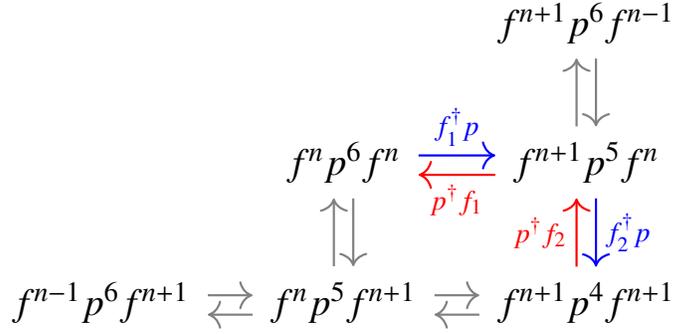

FIG. 3: Process 3 involves intermediate states with only a single hole on the oxygen site. Scales roughly as $\sim t^4(U_f + \Delta_{pf})^{-2}(2U_f + 2\Delta_{pf} + U_p)^{-1}$

The oxygen part is a slightly less trivial this time with $P_p p_\alpha^\dagger p_\beta^\dagger p_\mu p_\nu P_p = \delta_{\beta\mu}\delta_{\alpha\nu} - \delta_{\alpha\mu}\delta_{\beta\nu}$ We thus have

$$P_f \left( \sum_{\alpha\beta\mu\nu} f_{2\alpha} R_{f,2} f_{1\beta} R_{ff}^{(2)} f_{2\mu}^\dagger R_{f,1} f_{1\nu}^\dagger \left( [t_2 t_1^\dagger]^{\mu\beta} [t_1 t_2^\dagger]^{\nu\alpha} - [t_2 t_2^\dagger]^{\mu\alpha} [t_1 t_1^\dagger]^{\nu\beta} \right) \right) P_f. \tag{28}$$

Finally, for process 3, we can reuse all of the building blocks that we have put together so far. From Figure 3, we directly write the contribution

$$P \left[ p^\dagger t_1^\dagger f_1 \right] R_{f,1} \left[ p^\dagger t_2^\dagger f_2 \right] R_{ff}^{(2)} \left[ f_2^\dagger t_2 p \right] R_{f,1} \left[ f_1^\dagger t_1 p \right] P. \tag{29}$$

Following an identical argument to the previous section we arrive at

$$P_f \left( \sum_{\alpha\beta\mu\nu} f_{1\alpha} R_{f,1} f_{2\beta} R_{ff}^{(2)} f_{2\mu}^\dagger R_{f,1} f_{1\nu}^\dagger \left( [t_2 t_2^\dagger]^{\mu\beta} [t_1 t_1^\dagger]^{\nu\alpha} - [t_2 t_1^\dagger]^{\mu\alpha} [t_1 t_2^\dagger]^{\nu\beta} \right) \right) P_f, \tag{30}$$

by simply swapping the site 1 and site 2 labels in part of the exchange path.

## MULTIPOLE RANKS AND RESOLVENT OPERATORS

From the expressions above, we see that the interactions between the rare-earth ions can be represented by operators of the form $\sim f_{i,\alpha}^\dagger f_{i,\beta}$ mediated by the oxygens, intermingled with resolvents such as $R_{f,i}$, $R_{ff}^{(1)}$ and $R_{ff}^{(2)}$. As discussed in the main text, we consider the case where the resolvents do not contain the CEF potential and are thus afforded the full spherical symmetry of the free-ion. Because of this symmetry, we can write the resolvents in terms of spherically symmetric projectors into each set of degenerate levels. Expanding out these sums and re-arranging terms, we find that

8

contributions to each process take the form

$$\sim \left(f_{1\alpha}^\dagger O_1 f_{1\beta}\right)\left(f_{2\mu}^\dagger O_2 f_{2\nu}\right), \tag{31}$$

where $O_1$ and $O_2$ are operators acting on rare-earth sites 1 and 2 built from rank-0 spherically symmetric projectors. The same rank counting outlined in the main text within the charging approximation thus holds true; operators of the form $f_{i,\alpha}^\dagger O_i f_{i,\beta}$ involve multipoles of rank-7 or lower.

**CHARGING ENERGY APPROXIMATION**

Here we give the details that link the general results to the final expressions for the charging approximation given in the main text. Recall that in this approximation we postulate the energies of the $f^n$ states predominantly depend on the total electron number $n$. With this approximation we write

$$R_{f,i} \sim \sum_{a \in f^{n+1}} \frac{|a\rangle_i \langle a|_i}{U_f^+ + \Delta_{pf}} \equiv \frac{1 - P_f}{U_f^+ + \Delta_{pf}}, \tag{32}$$

since $E_{f,a} - E_{f,0} \sim U_f^+$ for all states $|a\rangle \in f^{n+1}$. Since the structure of the perturbation forces us to be in the $f^{n+1}$ state, we can extend the sum to all excited states and thus arrive at the projector $1 - P_f$. Applying the same approximation to the innermost resolvents, we find

$$R_{ff}^{(1)} \sim \frac{1 - P_f}{U_f^+ + U_f^-}, \qquad R_{ff}^{(2)} \sim \frac{1 - P_f}{2U_f^+ + U_p + 2\Delta_{pf}}, \tag{33}$$

where $U_f^-$ is the energy difference $E_{f,a} - E_{f,0}$ when $|a\rangle$ is a state in the $f^{n-1}$ manifold. We can now use these resolvents to simplify the contributions from the three processes (Figs. 1, 2, and 3) considered earlier, and shown in Eqs. 23, 28, and 30. In each expression the ground state projectors $P_f$ drop out, and we can write the three parts

$$1 : \frac{1}{(U_f^+ + \Delta_{pf})^2(U_f^+ + U_f^-)} P_f \left(\sum_{\alpha\beta\mu\nu} f_{1\alpha} f_{2\beta}^\dagger f_{2\mu} f_{1\nu}^\dagger \left[t_2 t_1^\dagger\right]^{\beta\alpha} \left[t_1 t_2^\dagger\right]^{\nu\mu}\right) P_f,$$

$$2 : \frac{1}{(U_f^+ + \Delta_{pf})^2(2U_f^+ + U_p + 2\Delta_{pf})} P_f \left(\sum_{\alpha\beta\mu\nu} f_{2\alpha} f_{1\beta} f_{2\mu}^\dagger f_{1\nu}^\dagger \left(\left[t_2 t_1^\dagger\right]^{\mu\beta}\left[t_1 t_2^\dagger\right]^{\nu\alpha} - \left[t_2 t_2^\dagger\right]^{\mu\alpha}\left[t_1 t_1^\dagger\right]^{\nu\beta}\right)\right) P_f,$$

$$3 : \frac{1}{(U_f^+ + \Delta_{pf})^2(2U_f^+ + U_p + 2\Delta_{pf})} P_f \left(\sum_{\alpha\beta\mu\nu} f_{1\alpha} f_{2\beta} f_{2\mu}^\dagger f_{1\nu}^\dagger \left(\left[t_2 t_2^\dagger\right]^{\mu\beta}\left[t_1 t_1^\dagger\right]^{\nu\alpha} - \left[t_2 t_1^\dagger\right]^{\mu\alpha}\left[t_1 t_2^\dagger\right]^{\nu\beta}\right)\right) P_f.$$

We can further massage the expressions for each process to put the $f$ operators into a standard order. Finally, the swapping of 1 and 2 generates an identical contribution and thus a factor of

two. Collecting the contributions from all three processes and dropping any single-site operators generated by commuting the pieces, we have the final result for $H_{\text{eff}}$

$$H_{\text{eff}} = \sum_{\alpha\beta\mu\nu} \left(P_{f,1} f_{1\alpha}^\dagger f_{1\beta} P_{f,1}\right) \mathcal{I}_{12}^{\alpha\beta\mu\nu} \left(P_{f,2} f_{2\mu}^\dagger f_{2\nu} P_{f,2}\right), \qquad (34)$$

where the interaction matrix is defined as

$$\mathcal{I}_{12}^{\alpha\beta\mu\nu} \equiv \frac{2}{(U_f^+ + \Delta_{pf})^2} \left[ \left( \frac{1}{U_f^+ + U_f^-} + \frac{2}{2U_f^+ + U_p + 2\Delta_{pf}} \right) [t_2 t_1^\dagger]^{\mu\beta} [t_1 t_2^\dagger]^{\alpha\nu} - \frac{2 \left[t_2 t_2^\dagger\right]^{\mu\nu} \left[t_1 t_1^\dagger\right]^{\alpha\beta}}{2U_f^+ + U_p + 2\Delta_{pf}} \right]. \qquad (35)$$

as given in the main text. An alternative route to the super-exchange interactions would be to integrate out the oxygen degrees of freedom first. This down-folding gives an effective $f$-$f$ hopping of order $t_{\text{eff}} \equiv t^2/\Delta_{pf}$. Within the approach discussed here, this corresponds to taking $\Delta_{pf} \gg U_f^\pm$, eliminating all but process 1 at leading order. This then takes the form of a second-order result in the effective hopping $t_{\text{eff}}$

**OTHER MICROSCOPIC INTERACTIONS**

Here we outline some other possible sources of interactions with the J = 15/2 and J = 8 manifolds of $Dy^{3+}$ and $Ho^{3+}$.

### Magnetic multipole interactions

Being of magnetostatic origin, the scale of the magnetic multipole interactions (MMI) are simple to address. These are only directly relevant for $Dy_2Ti_2O_7$, as the transverse couplings in non-Kramers doublet relevant for $Ho_2Ti_2O_7$ are time-reversal even. Generally, the scale of the interaction between a rank-$K$ and rank-$K'$ magnetic multipole is approximately given by [13]

$$\mathcal{M}_{KK'} \sim \chi_{K-1} \chi_{K'-1} \langle r^{K-1} \rangle \langle r^{K'-1} \rangle \left( \frac{\mu^{K+K'} \mu_0}{4\pi} \right) \frac{1}{R^{K+K'+1}}, \qquad (36)$$

where $\chi_K$ is the order $K$ Stevens' parameter [14], $\langle r^K \rangle$ is an atomic radial integral [15], $\mu$ is the magnetic moment of the ion and $R$ is the distance between the ions. This expression is somewhat conservative as we have assumed the multipole moments are maximal, including factors of $J^K$ and $J^{K'}$ in $\mu^{K+K'}$ where J = 15/2. Projecting into the ground state doublet will quench these moments and further reduce the estimate. This is especially true for $Dy_2Ti_2O_7$ where the ground state

doublet is primarily composed of $|\pm 15/2\rangle$ states and thus has predominantly rank-15 moments that could contribute the transverse couplings. Using $\mu \sim 10\,\mu_B$, and we find these MMI are small beyond the dipole-dipole term $\mathcal{M}_{11} \sim 1.4$ K. The more useful number is is $D \equiv 4\,(5\mathcal{M}_{11}/3) \sim 9$ K as discussed in the main text. For dipole-octupole coupling one finds $\mathcal{M}_{13} \sim -16$ mK while for octupole-octupole one finds $\mathcal{M}_{33} \sim 0.17$ mK. Given the factor of $R^{-(K+K'+1)}$, higher-rank interactions are quickly suppressed. The nearest-neighbor dipole-octupole coupling is largest and necessarily projects into exchanges of the form $\sim S_i^z S_j^x + S_i^x S_j^z$. Finally, one must take into account that when the octupole moments are projected into the ground state doublet, their contribution is further reduced. The suppression factors are $\sim \delta/\alpha$, which reduce the dipole-octupole terms by a factor of $\sim 0.1$ and the octupole-octupole terms by $\sim 0.01$. The induced transverse terms after rotation is thus of order $\lesssim 10^{-3}$ mK. The quantum effects are then completely negligible, and thus the MMI can be safely ignored beyond dipole-dipole interactions discussed in the main text.

**Electric multipole interactions**

Electric multipole interactions (EMI) are only directly relevant for $Ho_2Ti_2O_7$ as the transverse couplings in the Kramers doublet relevant for $Dy_2Ti_2O_7$ are time-reversal odd. Generally, the scale of the EMI between a rank-$K$ and rank-$K'$ electric multipole is approximately given by [16]

$$\mathcal{E}_{KK'} \sim \chi_K \chi_{K'} \langle r^K \rangle \langle r^{K'} \rangle \frac{\mathrm{J}^{K+K'}}{R^{K+K'+1}}\left(\frac{e^2}{4\pi\epsilon_0}\right), \tag{37}$$

where $e$ is the elementary charge, $\epsilon_0$ is permittivity of vacuum and the remainder of the notation is the same as in the previous section. We have included factors of $\mathrm{J}^K$ and $\mathrm{J}^{K'}$ to account for possible matrix element effects as conservatively as possible. The largest of these couplings is the electric quadrupole-quadrupole coupling at order $\mathcal{E}_{22} \sim 63$ mK. Due to the composition of the ground doublet, transverse couplings originating from EMI will carry additional factors of $\sim (\delta/\alpha)^2$, as in the MMI. The scale of these contributions are then $\lesssim 1$ mK. The higher order interactions are even smaller and so the electric multipolar interactions are not a significant source of transverse couplings in $Ho_2Ti_2O_7$.

**Lattice mediated interactions**

Additional interactions could be mediated through spin-phonon coupling [17]. Due to time-reversal invariance, these can only directly generate interactions between multipoles of even rank.

For $Dy_2Ti_2O_7$ these vanish under projection into the ground state CEF doublet, and thus do not contribute directly to the transverse terms. For $Ho_2Ti_2O_7$, such even-rank interaction can contribute directly to the transverse terms. To evaluate the plausibility of such a scenario, we consider two models of spin-phonon interaction, a site-phonon and bond-phonon model. In the site-phonon model, we consider only the coupling of the on-site multipoles to the local lattice distortion. In the same way as the CEF, when represented as Stevens' operator equivalents, these can only include operators up to rank-6. When the site-phonon is integrated out, these operators are squared so only interactions of rank-12 or less can be generated. In a bond-phonon type model, we imagine the lattice distortion modifying the exchanges already present in the system. Given the argument in the main text, we expect these to be predominantly rank-7 or lower and thus when the phonon is integrated out the induced interactions are rank-14 or smaller. Since these are not the direct rank-16 interactions needed to connect $|\pm 8\rangle$ to $|\mp 8\rangle$, they are suppressed by the sub-leading doublet components by same mechanisms discussed in the main text. Additionally, there will be a further suppression from the energy scale of the phonon that was integrated out. We thus can safely ignore the spin-lattice couplings in our analysis.

**Direct exchange**

We have accounted for some of the inter-site Coulomb interaction in our treatment of EMI. Other contributions such as direct exchange between the rare-earth ions is likely to be of the same order or smaller given their large separation. Generally, one could write pair-wise direct exchange contributions in the form

$$\sum_{ij} \sum_{\alpha\beta\mu\nu} \mathcal{U}_{ij}^{\alpha\beta\mu\nu} f_{i\alpha}^\dagger f_{i\beta} f_{j\mu}^\dagger f_{j\nu} \tag{38}$$

As we have seen, the one-electron form of the operators at each site restrict this to generate multipolar interactions of rank-7 or less. Thus we have the same suppression factor $(\delta/\alpha)^2$ as discussed in the main text. Given the scale of the EMI, we then expect these terms to be negligible. We note that direct exchange interactions between the rare-earth and oxygen sites could be present. However, due to the need to include additional electron transfers to generate interactions and the filled shell of the oxygen site, we would not expect these terms to be important.

12



\end{document}